\documentclass[aps,amsfonts,amsmath,prd,preprint,nofootinbib]{revtex4}
\usepackage{epsf,mathrsfs}
\usepackage{graphicx} 

\newcommand{\beq}{\begin{equation}}
\newcommand{\eeq}{\end{equation}}

\begin{document}

\title{Holographic description of vacuum bubbles}

\author{Jaume Garriga}
\address{
Departament de F{\'\i}sica Fonamental i \\Institut de Ci{\`e}ncies del Cosmos, 
Universitat de Barcelona,\\
Mart{\'\i}\ i Franqu{\`e}s 1, 08028 Barcelona, Spain}

\begin{abstract}

We discuss a holographic description of vacuum bubbles, with possible implications for a consistent description of the multiverse. In particular, we elaborate on the recent observation by Maldacena, that the interior of AdS bubbles can be described in terms of CFT degrees of freedom living on the worldsheet of the bubble wall. We consider the scattering of bulk gravitons in the ambient parent vacuum, off the bubble wall. In the dual description, the transmission coefficient is interpreted as the probability that a graviton is absorbed by the worldsheet CFT degrees of freedom. The result is in agreement with intuitive expectations. Conformal invariance is not exact in this setup, and the leading corrections due to the IR and UV cut-offs are displayed. Aside from bulk scattering states, we find that when a bubble nucleates within a parent dS vacuum, there is a zero mode of the graviton which describes lower dimensional gravity with a finite Newton's constant. This massless graviton lives within one Hubble radius away from the bubble wall. 
Possible implications for a fully holographic description of the inflating multiverse are briefly discussed.

\end{abstract}

\maketitle

\section{Introduction}

Transitions between inflating $dS$ vacuua and neighboring vacuua lead to the picture of a multiverse where regions with different low energy physics properties can be found sitting next to each other. The transitions may occur by quantum nucleation of localized bubbles of the new vacuua, within the inflating parent vacuum (an alternative mechanism is quantum diffusion, which we shall not consider here.)

The physics leading to nucleation, growth, and possible collision of bubbles is by now  reasonably well understood. However, the global properties of the eternally inflating multiverse have resisted, so far, a comprehensive description with a clear interpretation. An important missing ingredient is the so-called ``probability measure", which would allow us to make predictions for observations within the multiverse.

A new approach for handling these problems was recently proposed in \cite{GV1} and \cite{GV2} (see also \cite{F,BFLR}). The idea was to replace the $D$ dimensional space-time ``bulk" description of the multiverse, with a purely spatial theory without gravity at its $D-1$ dimensional future boundary. There, the degrees of freedom of a putative CFT would interact with 
$D-2$ dimensional defects, representing the future boundary of bulk domain walls separating the different vacua.
Precursors of this approach are Strominger's $dS/CFT$ correspondence \cite{dS}, and Susskind's dual ``Census taker's Hat"  \cite{FSSY,hat}. The latter involves a theory at the $D-2$ future boundary of a single domain wall, containing a Minkowski vacuum embedded in the inflating multiverse.

A significant obstacle in developing the proposal of Refs. \cite{GV1,GV2} was the description of $AdS$ bubbles. These contain a space-like singularity at the future boundary of their interior, and it was not clear what to do with it in the dual description. A related problem was the description of the (mostly null) future boundary of Minkowski bubbles. Based on the approach of  \cite{hat}, it was suggested in  \cite{GV1,GV2} that the future boundaries of Minkowski and $AdS$ bubble interiors should be simply removed from the $D-1$ regular part of the future boundary, and replaced with $D-2$ theories at the edges of the corresponding ``holes". This heuristic prescription was adopted in order to implement a specific probability measure for the multiverse (based on a Wilsonian cut-off in the dual theory). However, the justification for such prescription awaits a better understanding of $AdS$ bubbles (and of the relation between the Census Taker's Hat approach of \cite{hat} and the approach of Refs. \cite{GV1,GV2}). 
A recent observation by Maldacena \cite{M} may help bringing this understanding a tad closer.

In Ref. \cite{M}, Maldacena points out that the interior of $AdS$ bubbles nucleating in a Minkowski (or in a $dS$) vacuum, can be described in terms of 
CFT degrees of freedom living on the worldsheet of the domain wall. The range of the holographic coordinate is limited, and so the CFT has  a UV cut-off, corresponding to the $AdS$ radius $\ell$ at the location of the domain wall $\lambda_{UV}=\ell^{-1}$. Since the worldsheet has the internal geometry of $dS_n$ (with $n=D-1$), conformal invariance is also broken at an IR cut-off $\lambda_{IR}=a_w^{-1}$ corresponding to the intrinsic worldsheet curvature radius $a_w$. 

The device of substituting the bubble interior with a worldsheet field theory seems to allow for a regular description of $AdS$ bubbles, bypassing the need to consider singularities at the future boundary.
The purpose of this paper is to elaborate on the proposal of Maldacena, by considering the interaction of bulk gravitons in the ambient parent vacuum with the CFT degrees of freedom. This step may help 
clarifying whether the bulk and worldsheet degrees of freedom might fit together in a fully holographic description of the multiverse.

As we shall see, the case of bubbles nucleating in an inflating vacuuum is somewhat different from that of bubbles nucleating in a Minkowski vacuum. In the case of an 
inflating parent vacuum, the worldsheet field theory contains lower ($n=D-1$) dimensional gravity. This suggests the possibility of a second duality, where the wordsheet degrees of freedom may be represented at the future boundary in terms of a theory without gravity.

The paper is organized as follows. In Section I we consider the background solutions. In Section II, we set up the description of linearized gravitons in the bubble background.
Section III discusses the worldsheet reduction of the bulk graviton dynamics, and the presence of a graviton zero mode. Section IV discusses the interaction of bulk graviton scattering states with the domain wall, both in the bulk D dimensional picture and in the dual worldsheet CFT picture. The transmission coefficient of gravitons accross the wall has a dual interpretation as the absorption of bulk gravitons by the CFT degrees of freedom. Section V summarizes our conclusions.

\section{Background solutions}

Consider a scalar field $\phi$ with potential $V(\phi)$ in $D=n+1$ dimensions. Let us assume that the potential has two local minima, with vacuum energies 
$V_1$ and $V_2$, separated by a domain wall.
Coupling the scalar field to standard Einstein gravity, and assuming maximal symmetry for the solution connecting
the two vacua, the metric will be given by 
\begin{equation}
ds^2 = a^2(z) (dz^2 + \gamma_{\mu\nu} dx^{\mu}dx^{\nu}).
\end{equation}
The case of primary interest for us is that of bubbles of a given vacuum embedded in the other vacuum. In this case,  $\gamma_{\mu\nu}$ is the metric of n-dimensional
de Sitter space ($dS_n$) of unit radius. For generality, the cases where $\gamma_{\mu\nu}$ is Minkowski ($M_n$) or Anti-de Sitter ($AdS_n$) are also 
considered. These may also be of interest depending on the values of the vacuum energies and the tension of the domain wall separating 
them (see below). 

The scalar field equation
\begin{equation}
\phi''+(n-1){\cal H} \phi' - a^2 {dV\over d\phi} = 0, \label{field}
 \end{equation}
 has a solution $\phi=\phi_w(z)$ which interpolates between the two local minima. Here, we have introduced ${\cal H} = a'/a$, 
and primes denote derivatives with respect to 
 $z$.
 In the thin wall case, the gradient of $\phi$ is concentrated near some value of the conformal coordinate $z=z_w$. In this case the worldsheet geometry of the bubble wall is that of $dS_n$ of radius $a_w=a(z_w)$.
Note that $z$ is a space-like coordinate, and the warp factor $a(z)$ obeys a ``rotated" Friedmann equation
\begin{equation}
{\cal H}^2 = -K + {16 \pi G_D \over n (n-1)} a^2 p_z, \label{Friedmann}
\end{equation}
where 
\begin{equation}
p_z = {\phi'^2\over 2 a^2} - V(\phi).
\end{equation}
$K=-1$ corresponds to the $dS_n$ foliaton, which as mentioned above is the case relevant for false vacuum decay. 
Nonetheless, we keep $K$ explicit in order to include more general solutions,
without spatial slicings where the domain wall is compact. When $V_1< 0$ and $V_2 \leq 0$, the two vacua can be 
separated by a flat domain wall ($K=0$) provided that this has a precisely tuned critical tension (as happens for certain BPS solutions).
Explicit expressions for the critical tension $T_{crit}$ will be given below in the thin wall limit. 
When both $V_1$ and $V_2$ are strictly negative, then the geometry of the worldsheet will be 
$AdS_n$ if the tension is lower than the critical tension. This case corresponds to $K=1$. 
\begin{figure}
  \begin{center}
 \includegraphics[width= 20 cm]{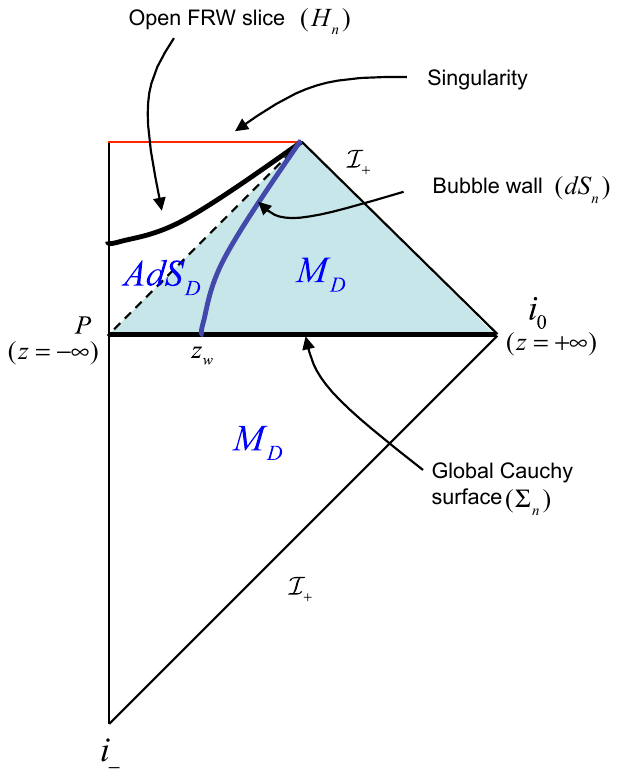}
  \end{center}
  \caption{Causal diagram representing the nucleation of an $AdS_D$ bubble in a Minkowski $M_D$ vacuum. $z$ is the holographic conformal coordinate,
  orthogonal to the $dS_n$ foliation. The bubble wall is at the fixed value $z=z_w$.}
   \label{MAdSl}
\end{figure}

\begin{figure}
  \begin{center}
 \includegraphics[width= 20 cm]{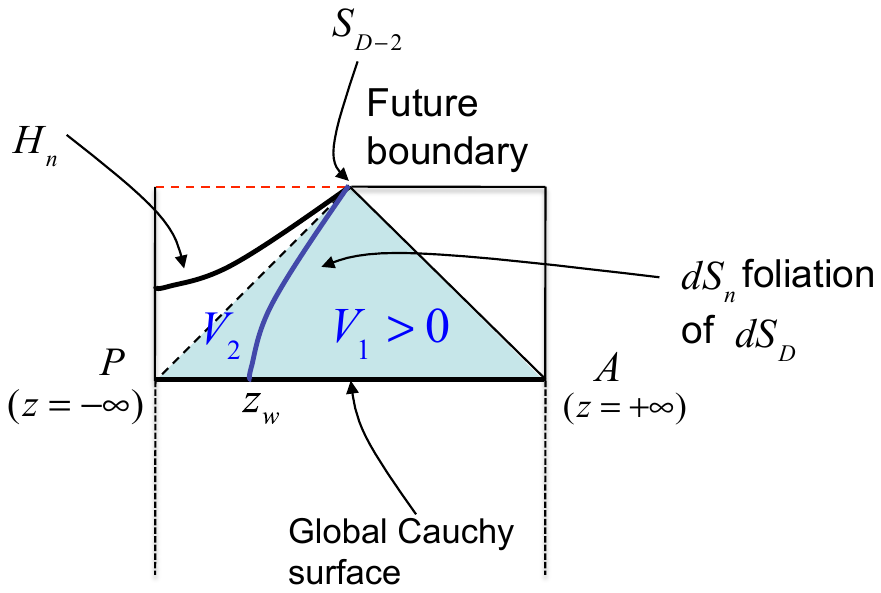}
  \end{center}
  \caption{Causal diagram representing the nucleation of a bubble with vacuum energy $V_2$ inside of a $dS_D$ vacuum with vacuum energy 
  $V_1>0$. The $dS_n$ foliation parallel to the worldsheet of the bubble wall covers the shaded region of the diagram, but not the interiors of the 
  future lightcones from the center of symmetry of the bubble, $P$, or of its antipodal point $A$. The dotted line at the future boundary corresponds 
  to a singularity, a regular horizontal boundary, or a null ``Minkowski" hat, depending on whether $V_2$ is negative, positive or zero. The vertical lines
  represent the worldlines of the north and south poles of a compact foliation, to which the global cauchy surface drawn in the figure belongs.
}
   \label{dSAdS}
\end{figure}

\section{perturbations}

Let us now consider the perturbed ansatz
\begin{equation}
ds^2 = a^2(z) [dz^2 + (\gamma_{\mu\nu}+h_{\mu\nu}) dx^{\mu}dx^{\nu}], \label{ansatz1}
\end{equation} 
where $h_{\mu\nu}(z, x)= h(z) Y_{\mu\nu}(x)$ is transverse and traceless
\begin{equation}
Y^{\mu}_{\mu} = Y^{\mu\nu}_{\ \ |\nu} = 0. \label{ansatz2}
\end{equation}
Here, the vertical stroke denotes covariant derivative with respect to the metric 
$\gamma_{\mu\nu}$, which is also used in order to raise and lower greek indices. 
The linearized Einstein's equations reduce to
\begin{equation}
h''+(n-1) {\cal H} h' + m^2 h =0,   \label{radial}
\end{equation}
and
\begin{equation}
(\Box_\gamma +2K- m^2) Y_{\mu\nu} = 0. \label{massivegraviton}
\end{equation}
Here $m^2$ is a separation constant, and the d'Alembertian $\Box_\gamma$ associated to the metric $\gamma_{\mu\nu}$ 
acts on the second rank tensor 
$Y_{\mu\nu}$ covariantly.  The preceeding equations can be obtained from the equations for tensor 
cosmological perturbations in a FRW universe with spatial curvature $K$ (see e.g. \cite{KS}).

In general, a solution of the linearized Einstein's equations with scalar matter 
{\em cannot} be brought to the form (\ref{ansatz1}) with (\ref{ansatz2}) . 
However, it is possible to write it in this way if the scalar field is unperturbed. This will be sufficient for 
our purposes in the case where the domain wall separating the two vacuum phases is very thin 
compared with the worldsheet curvature scale (For $D=4$, this follows from the analysis in \cite{GMST1,GMST2}. See \cite{IT} for a discussion of 
the general case).

Introducing
\begin{equation}
W = {n-1 \over 2} {\cal H},
\end{equation}
and the rescaled variable $\hat h = a^{n-1\over 2} h$, the radial equation (\ref{radial}) can be
written as
\begin{equation}
-\hat h'' + [W^2 + W'] \hat h = m^2 \hat h.  \label{Schrodinger}
\end{equation}
This describes the scattering of bulk gravitons off the domain wall separating the two vacua. 
In addition, when the spatial geometry is compact (as is the case, for instance when we consider the decay of a de Sitter vacuum,
with $V_1>0$), there is a normalizable bound state which carries the degrees of freedom of lower n-dimensional
gravity, as we now describe.

\section{Worldsheet reduction}

The Schrodinger equation (\ref{Schrodinger}) can be written in the language of supersymmetric quantum mechanics as:
\begin{equation}
S^\dagger S\ \hat h = m^2 \hat h, \label{susy}
\end{equation}
where we have introduced $S=-\partial_z +W$ and $S^\dagger=\partial_z + W$. The wave function
\begin{equation}
\hat h_0 \propto a^{n-1\over 2}(z), \label{zero}
\end{equation}
satisfies $S \hat h_0 = 0$. Clearly, $\hat h_0$ has no nodes, since $a(z)$ is positive definite, and hence it will be the ground state of
$S^\dagger S$ provided that it is normalizable. The corresponding eigenvalue is $m^2=0$.

For this mode, the metric perturbation 
\begin{equation}
h_{\mu\nu}(z,x) = h_0(z) Y_{\mu\nu}(x) = a^{-{n-1\over 2}} \hat h_0(z) Y_{\mu\nu}(x) \propto Y_{\mu\nu}(x),
\end{equation}
is independent of $z$, and in this sense we can think of it as a lower dimensional graviton which lives in 
the direction parallel to the world-sheet. Note that the perturbed metric takes the form
\begin{equation}
ds^2 = a^2(z) [dz^2 + g_{\mu\nu}(x) dx^{\mu}dx^{\nu}], \label{ansatz1}
\end{equation} 
where now $a_r^2 g_{\mu\nu}(x)$ is a vacuum solution of the n-dimensional Einstein's equations, with a lower dimensional cosmological constant 
$\Lambda_n = -[n(n-1)/2] K a_r^{-2}$. 
Here, $a_r=a(z_r)$ is the warp factor at some reference value of $z$. This value is conventional, but it makes sense to
take it near the peak of the graviton zero mode, where the lower dimensional gravity is localized.
Dimensionally reducing in the direction transverse to the worldsheet, the lower dimensional Newton's constant
$G_n$ is given in terms of the bulk D dimensional one by
\begin{equation}
{1\over G_n} = {a_r^{2-n}\over G_D} \int_{-\infty}^{\infty} dz\ a^{n-1}(z).\label{gn}
\end{equation}
Hence, $G_n$ will be non-vanishing whenever the zero mode (\ref{zero}) is square integrable.

Eq. (\ref{massivegraviton}), together with the fact that the graviton zero mode has eigenvalue $m^2=0$, suggests a KK-type reduction along the
$z$ direction, where the KK gravitons have squared masses given by $m^2$ (in units of the warp factor at the location of the reference worldsheet, $a_r$).
In the case where $\hat h_0$ is normalizable, the ground state is in the discrete spectrum. 
In general, there may be additional discrete modes, as well as the continuum. 

It is easy to show that in $D=n+1=4$ dimensions, the zero mode $\hat h_0$ is the only discrete one. For that, we consider the "supersymmetric" partner of 
(\ref{susy}),
\begin{equation}
SS^\dagger\ \tilde h = -\tilde h'' + [W^2 - W'] \tilde h = m^2 \tilde h, \label{susy2}
\end{equation}
where $\tilde h = S \hat h$. This can be obtained from (\ref{susy}) by applying $S$ on both sides of the equation. For the zero mode, the partner
vanishes identically, $\tilde h_0 =0$. Hence, the spectra of the Schrodinger equations (\ref{susy}) and (\ref{susy2}) will be the same except for 
the ground state of the first, which is missing in the second. Using the background equations of motion (\ref{field}) and (\ref{Friedmann}), 
the new potential is given by:
\begin{equation}
W^2-W'=4\pi G_D \phi'^2 - K. \quad (n=3). \label{n=3}
\end{equation}
Up to a constant, this is positive definite and therefore there are no additional states in the discrete spectrum. 
On the other hand, these may be present for $n>3$  (see e.g. \cite{dSdS}.\footnote{Ref. \cite{dSdS} introduces the $dS/dS$ correspondence, which has
significant overlap with the subject of the present paper.})

The continuum of KK modes corresponds to bulk gravitons which scatter off the bubble wall. 
If the warp factor $a(z)$ resembles that of $AdS_D$ for a range of $z$ 
spanning many $AdS$ radii, then the effect of the bulk KK modes in that region on the lower dimensional projection can be interpreted (in a certain energy range) as due to a CFT living in the dimensionally reduced space, and coupled to the zero mode graviton with strength $G_n$.

\subsection{Randall-Sundrum brane-world}

A particularly interesting case is that of a flat domain wall of critical tension (so that $K=0$) 
separating two degenerate $AdS_D$ vacua (i.e. local minima with negative energy density, $V_1=V_2<0$). In the thin wall limit, the warp factor is given
by $a(z) = \ell(|z|+1)^{-1}$, where $z=0$ is the location of the domain wall and $\ell$ is the $AdS_D$ radius. This is the well known Randall-Sundrum 
(RS) solution \cite{RS}.
In the dimensionally reduced picture, n-dimensional gravity is carried by the zero mode. On the other hand, the dynamics of the bulk KK modes 
has the same physical effects as a worldsheet $n-$dimensional  CFT, with a UV cut-off lengthscale of order $\ell$ on the brane.

If the tension of the brane separating the two degenerate vacua is larger than critical, then the RS flat solution generalizes to a $dS_n$ brane-world
($K=-1$). In the CFT description, in addition to the UV cut-off we now have an IR cut-off of the order of the $dS_n$ worldsheet radius. 
For instance, from (\ref{n=3}), the continuum of KK modes starts after the mass gap $m^2=-K=1$ \cite{GS}. This reflects the 
fact that conformal invariance of the CFT is broken by the background $dS_n$ geometry in which it lives.

\subsection{Bubble solutions}

It is easy to show that $\hat h_0$ is normalizable for bubble solutions ($K=-1$) provided that the background solution admits compact spatial slices.
This will always be the case for $V_1>0$ and arbitrary $V_2$, or when the tension of the bubble wall exceeds the critical tension for $V_1,V_2 \leq 0$. In this case the warp factor $a(z)\propto e^{-|z|}$ at $z\to \pm \infty$. Geometrically, these two limits 
correspond to the two "antipodal" fixed points of $dS_n$ symmetry, i.e. the points $P$ and $A$ in Fig. 2.

\section{Scattering of KK gravitons}

Maldacena has recently suggested that the description of the interior of AdS vacuum bubbles embedded in a Minkowski or in a $dS$ vacuum 
can be done in terms of a CFT living on the bubble wall, provided that the size of the bubble is much bigger than the $AdS$ radius. The situation is quite similar to the case of the $dS_n$ Randall Sundrum brane world discussed above.  The CFT will have an IR cut-off corresponding to the radius $a_w$ of the $dS_n$ worldsheet, and a UV cut-off corresponding to the $AdS_D$ 
radius $\ell$.

In this dual picture, there is no bubble interior. Gravitons impinging from the parent vacuum on the brane would reflect completely off the bubble wall 
if it wasn't for the fact that the CFT degrees of freedom can absorb them. Clearly, the probability of transmission across the wall in the 
bulk picture corresponds to the absorption probability in the dual picture.

\subsection{Bulk calculation (in $D=n+1=4$)}

In the thin wall limit, $\phi'^2= a_w T \delta(z-z_w)$. Here, $a_w$ is the intrinsic radius of the wall, and $T$ is the wall tension. The scattering potential reduces to
\begin{equation} 
-\tilde h'' + 2 A a_w \delta(z-z_w)\  \tilde h = k^2 \tilde h,
\end{equation}
where
\begin{equation}
A= 2 \pi G_4 T 
\end{equation}
and 
\begin{equation}
k^2 = m^2+K.  \label{gap}
\end{equation}
The transmission probability for scattering states of the form $e^{{\pm} i k z}$  incident from either side of the barrier 
is given by
\begin{equation}
|{\cal T}|^2 = {k^2 \over k^2 + (a_w A)^2} = {k_w^2 \over k_w^2 + (2\pi G_4 T)^2}.\label{absorption}
\end{equation}
In the second equality, we have introduced $k_w=k/a_w$, which is the physical wavelength of the gravitons at the location of the wall.
Physically, the gravitational field of a domain wall of tension $T$ is characterized by the lengthscale $(G_4 T)^{-1}$. Hence, it is not
surprising that gravitons with $k_w^2 \gg (G_4 T)^2$ are transmitted through it without much impediment. 
For $k_w^2 \ll (G_4 T)^2$, the transmission probability is inversely proportional to the square of the wall tension $|{\cal T}|^2 \sim (k_w/G_4 T)^2$.

\subsection{Dual interpretation}

In the dual picture, $|{\cal T}|^2$  is the probability for the graviton to be absorbed by the CFT degrees on the brane.

\subsubsection{Flat walls ($K=0$).}

To begin with, let us consider the particular case where a region of AdS, with curvature radius $\ell$, is separated from a region of Minkowski space, by a flat domain wall of critical 
tension
\begin{equation}
T_{crit} = {1\over 4\pi G_4 \ell}.
\end{equation}
In this case, for $k_w \ll \ell^{-1}$ the absorption coefficient is given by
\begin{equation}
|{\cal T}|^2 \sim \ell^2 k_w^2  = c\ (G_4 E^2). \label{absorptionflat}
\end{equation}
In the second equality, $E=k_w$ is the energy of the graviton incident
from the Minkowski side of the wall. The AdS side is removed from the picture and substituted with 
\begin{equation}
c \sim \ell^2/G_4 
\end{equation}
CFT degrees of freedom
living on the wall. The result (\ref{absorptionflat}) is not surprising. The first factor $c$ accounts for the number of species
which interact with the incident graviton, while the second is the cross section for gravitational interaction per species. For 
$k_w \gg \ell^{-1}$, we are at energies above the UV cut-off scale, where the CFT interpretation is no longer valid. Note that at energies comparable
to the cut-off scale,  absorption saturates the unitarity bound (the exact expression for the transmission coefficient $|{\cal T}|^2$ is of course unitary, 
but the dual interpretation in terms of a conformal theory should not  be extrapolated above the UV cut-off scale).

\subsubsection{Bubbles (K=-1)}

Let us now consider the case of $AdS_4$ bubbles embedded in Minkowski space. In this case the tension (which follows from Israel's matching conditions) 
is lower than the critical tension:
\begin{equation}
T= T_{crit} \left[ \sqrt{1+(\ell/a_w)^2} -(\ell/a_w)\right] . \label{tension}
\end{equation}
From (\ref{gap}) the energy of the gravitons at the location of the wall is given by
\begin{equation}
E^2= m^2 a_w^{-2} = k_w^2+(1/a_w)^2. \label{gap2}
\end{equation}
Note that the minimum energy for a scattering state of the graviton is larger than the infrared mass gap 
\begin{equation}
E > \lambda_{IR} \equiv a_w^{-1}.
\end{equation}
If the incident graviton has exactly this energy, then the absorption coefficient (\ref{absorption}) is zero, indicating that there is no phase space in the CFT for the 
graviton to decay. Note also that the infrared mass gap is of the order of the intrinsic Gibbons-Hawking temperature of the worldsheet CFT:
\begin{equation}
T_{CFT}={1\over 2\pi a_W} \equiv {\lambda_{IR}\over 2\pi}. \label{gh}
\end{equation}
 Substituting (\ref{tension}) in (\ref{absorption}) we have 
\begin{equation}
|{\cal T}|^2={(E^2-\lambda_{IR}^2) \over 
E^2-\lambda_{IR}^2 +
(\lambda_{UV}/2)^2[\sqrt{1+(\lambda_{IR}/\lambda_{UV})^2}-(\lambda_{IR}/\lambda_{UV})]^2}.
\end{equation}
In the range $\lambda_{IR}\ll E\ll \lambda_{UV}$, where the CFT interpretation is supposed to hold, the dominant term is of course
the same as in (\ref{absorptionflat}). Nonetheless, it may be of interest to look at the leading IR and UV  corrections.

Assume first that the energy is closer to the infrared than to the UV cut-off 
\begin{equation}
\lambda_{IR} < E \ll (\lambda_{IR}\lambda_{UV})^{1/2}. \label{irange}
\end{equation}
Then, $(\lambda_{IR}/E)^2 \gg (\lambda_{IR}/\lambda_{UV})$, and the leading correction takes the form \footnote{We thank Sugumi Kanno for pointing out an error in a previous version of the manuscript,
where a subleading correction was included in the expansion, smaller than $\lambda_{IR}/\lambda_{UV}$.}
\begin{equation}
|{\cal T}|^2 = c (G_4 \Delta E^2) [1+O(\lambda_{IR}/\lambda_{UV})]
\end{equation}
where $c \equiv 4\ell^2/G_4$ and $\Delta E^2 = E^2-\lambda_{IR}^2$. Note that in the range (\ref{irange}), we have $c(G_4\Delta E^2)\ll 1$. The absorption coefficient is smaller in this range than for flat walls, and, as mentioned above, it goes to zero as the energy approaches the IR cut-off.

Consider now the interval where 
\begin{equation}
(\lambda_{IR}\lambda_{UV})^{1/2} \ll E. 
\end{equation}
In this case $(\lambda_{IR}/E)^2 \ll (\lambda_{IR}/\lambda_{UV})$, and we have 
\begin{equation}
|{\cal T}|^2 = {c (G_4 E^2) \over 1+ c (G_4 E^2)} \left[ 1+{2\over1+ c (G_4 E^2)}{\lambda_{IR}\over \lambda_{UV}}+ O(\lambda_{IR}^2/E^2) \right].
\end{equation}
The correction linear in the ratio of cut-off scales is intriguing. In principle, an enhancement of the absorption rate might be expected 
due to stimulated decay of the KK graviton at the finite Gibbons-Hawking temperature (\ref{gh}) of the worldsheet CFT. However, we have not been able to find a good reason why this should be linear in $T_{CFT}$.

\section{Summary  and outlook}

We have explored the point of view where the interior of $AdS_D$ bubbles can be described in terms of a CFT living on the $dS_n$ worldsheet of the bubble wall ($D=n-1$).  Here, we summarize our findings and discuss some possible implications.

\subsection{Graviton absorption by the CFT}

In $D=4$, we have calculated the absorption probability for gravitons in the ambient parent vacuum to decay into the CFT degrees of freedom at the moment when they hit the bubble wall. The result is in agreement with intuitive expectations. The leading correction to the absorption rate has an intriguing linear dependence on the ratio of IR to UV cut-off scales
. This may be due to stimulated absorption by the CFT at the intrinsic Gibbons-Hawking temperature of the worldsheet. A more thorough discussion of such corrections in arbitrary dimension will be presented elsewhere.

\subsection{Complementarity}

Maldacena has suggested that the whole space-time region inside of an $AdS_D$ bubble can be described in terms of the boundary CFT living on 
the worldsheet of the bubble wall. This includes the the singularity which develops at the future boundary of the bubble interior. It would therefore be interesting if this could be removed from the picture and substituted with a regular description in terms of a worldsheet CFT.

In view of the preceeding analysis, a perhaps more precise statement seems to be the following. In the case of $AdS_n$ bubbles nucleating in Minkowski space, 
the woldsheet CFT correctly describes the effect of the bubble on the remaining Minkowski space at energies smaller than the UV cut-off, $E\ll \lambda_{UV}$. At energies comparable or above the UV cut-off, 
the wordsheet theory is no longer conformal, but we may perhaps adopt the point of view that there is still an effective theory on the worldsheet which accounts for the bulk modes inside the bubble. This would 
correspond to integrating the KK modes to the interior of the worldsheet in the dS foliation (that is, the modes in the shaded region in Fig. 1, to the left of the worldsheet).
Knowledge of the quantum state in the dual description should then be sufficient in order to determine the quantum state on an early slice $H_n$ of the open FRW universe within the bubble (see Fig 1). 

This aspect of complementarity is quite reasonable, and conforms with standard practice in explicit calculations of the quantum state of linearized cosmological perturbations in the ``one bubble" open universe (see e.g. \cite{GMST1,GMST2}). Indeed, the quantization is done on a global cauchy surface, such as the one labeled by $\Sigma_n$ in Figs. 1 and 2, which cuts through the region covered by the $dS_n$ slicing of the parent vacuum. The wave function is then propagated into the FRW patch, which is covered by the hyperbolic slicing $H_n$. Presumably, any quantum state in the dual picture be translated into a quantum state of bulk fields on the global Cauchy surface, and then propagated into $H_n$ (which lies in its future development).

Knowledge of the quantum state on $H_n$  allows for semiclassical probabilistic predictions inside of this FRW universe (such as, for instance, the number density of galaxies present at any given time). Unfortunately, at the singularity, the semiclassical description breaks down, and in the absence of a good ``dictionary", the worldsheet description does not seem to tell us much about what will happen near the big crunch, or whether there will be a bounce.
 
Let us now elaborate more precisely on the notion of complementarity. First, it should be stressed that the dual worldsheet theory plus bulk modes outside the bubble will always be in a pure state if the initial condition is a pure state (such as a $dS_n$ invariant state, plus possible excitations which may be included as initial conditions).  

Now, suppose for instance, that a radioactive atom falls inside of an $AdS$ bubble. The worldsheet theory will certainly learn about this, as the atom crosses the bubble wall. The atom will then decay at some  value of FRW time $t_d$, and the probability distribution for $t_d$ can be calculated from the boundary theory. Nonetheless, it is not so clear whether the decay of a particular atom will be ``communicated" to the boundary theory, whose wave function initially included all possibilities for $t_d$.   

Note that, if the wave function for a particular cosmological mode on the FRW slicing happens to collapse (say, by the effect of a measurement) to  a particular amplitude $a_k$, this would seem to force the collapse of the wave function in the dual description, so that the two unitary descriptions following the collapse are equivalent.

A somewhat different perspective arises  if we adopt the point of view that the collapse of the wave function is a fiction, and that all semiclassical branches of it are described by the pure state of the dual picture. In this case, the disintegration of the atom in a particular realization of its infall into the bubble need not be communicated to the boundary theory, which describes ``live" and ``dead" atoms at all times. This interpretation (which may be referred to as weak complementarity) seems more plausible to us than the possibility discussed in the preceding paragraph. 

\subsection{Bubbles nucleating in an inflating background}

For bubbles nucleating in a $dS_D$ background, we have seen that in addition to the graviton scattering modes, there is a normalizable graviton zero mode, bound within the spacetime region covered by the $dS_n$ foliation parallel to the worldsheet. The zero mode is not necessarily peaked on the bubble wall, but it will always have some overlap with the inside of the bubble. In the dual picture this will mediate a lower dimensional gravitational interaction amongst the worldsheet degrees of freedom.

This suggests that, for bubbles nucleating in a $dS_D$ parent vacuum, we may benefit from taking a further step (see also \cite{dSdS}).
Instead of just integrating out the KK gravitons on the $AdS_D$ side of the wall (in order to obtain a worldsheet CFT), we may integrate out the whole region covered by the worldsheet $dS_n $ foliation (shaded region in Fig. 2). This would define a putative field theory (which we may denote as $FT_n$ for the purposes of the present discussion) on the lower dimensional $dS_n$. The degrees of freedom of $FT_n$ are coupled to the lower dimensional gravity carried by the zero mode, with coupling $G_n$ given in Eq.  (\ref{gn}).

Irrespective of the sign of the vacuum energy inside of the new bubble, this worldsheet theory can be used in order to describe the effect of the nucleated bubble on the remainder of the inflating space (in  the causal diagram of Fig. 2, this is the region contained in the future light-cone from the point $A$ antipodal to $P$). 

The worldsheet theory may also be useful for calculating the probabilities for events inside of the future light cone from $P$, as we described in the previous subsection. However, the situation seems to be different depending on the nature of the new vacuum inside the bubble. 
For Minkowski and $AdS$ bubbles, all interesting physics happens early on after nucleation, and probabilistic predictions in the semiclassical regime will depend strongly on the initial conditions at $H_n$, which are in turn determined by the dual worldsheet theory $FT_n$. 
On the other hand, if the new vacuum is also $dS$, this may drive eternal inflation on its own, and the probabilities for
events happening at late times after nucleation bear little memory of the quantum state at the initial surface $H_n$. In this sense, $FT_n$ may not necessarily be too useful for the description of the interior of such bubbles. 

\subsection{Implications for holography at the future boundary}

As we just described, for bubbles nucleating in an ambient inflating vacuum, the worldsheet field theory $FT_n$ on $dS_n$ is coupled to lower $n$ dimensional gravity. 
Hence, we may be entitled to a further holographic description of this $n$-dimensional theory in terms of a theory without gravity at its future boundary. 
Note that the future boundary of the whole $dS_n$ foliation parallel to the worldsheet converges on a $D-2$ sphere, embedded in the $D-1$ 
dimensional future boundary of the bulk $D$ dimensional spacetime (see Fig. 2). Hence, the mapping of $FT_n$ onto a theory at the future boundary 
of the $dS_n$ foliation is quite natural. Although $FT_n$ is by no means conformal, we may still hope for a dual description of it in terms of an $n-1$ dimensional 
CFT on the $D-2$ sphere. 

As mentioned in the Introduction, a proposal for a dual description of the inflating multiverse in terms of a CFT at the future boundary was presented in \cite{GV1,GV2}. In this context, the description of the interiors of Minkowski and $AdS$ bubbles was left as an open issue, while the effects of such bubbles on the rest of space-time were ignored. The present approach may be useful for completing a consistent boundary description of the inflating multiverse.

\subsection*{Note Added:}

While this work was being prepared, an interesting paper by Harlow and Susskind \cite{HS} appeared, which substantially overlaps with the present one.


\section{Acknowledgements}

It is a pleasure to thank the organizers of the YKIS10 workshop at the Yukawa Institue for hospitality. 
I thank Roberto Emparan, Tomeu Fiol, Sugumi Kanno, David Mateos and Yuko Urakawa for 
very useful conversations. Special thanks are due to Alex Vilenkin for motivating discussions.
This work was supported in part by the Fundamental Questions Institute, and by grants DURSI 2009 SGR 168, 
MEC FPA 2007-66665-C02 and CPAN CSD2007-00042 Consolider-Ingenio 2010

\end{document}